\documentclass[twocolumn,showpacs,preprintnumbers,amsmath,amssymb]{revtex4}

\usepackage{graphicx}
\usepackage{dcolumn}
\usepackage{bm}

\begin{document}

\title{Simple Estimation of $X^{ - }$ Trion Binding Energy in Semiconductor Quantum Wells}

\author{R.~A.~Sergeev$^{1}$}
\author{R.~A.~Suris$^{1}$}
\author{G.~V.~Astakhov$^{1,2}$}
\author{W.~Ossau$^{2}$}
\author{D.~R.~Yakovlev$^{1,3}$}
\affiliation{
$^{1}$A.F.Ioffe Physico-Technical Institute, Russian
Academy of Sciences, 194021, St.Petersburg, Russia \\
$^{2}$Physikalisches Institut der Universit\"{a}t
W\"{u}rzburg, 97074 W\"{u}rzburg, Germany \\
$^{3}$Experimentelle Physik 2, Universit\"{a}t Dortmund, 44221
Dortmund, Germany}

\date{\today}

\begin{abstract}
A simple illustrative wave function with only three variational
parameters is suggested to calculate the binding energy of
negatively charged excitons ($X^{-}$) as a function of quantum
well width. The results of calculations are in agreement with
experimental data for GaAs, CdTe and ZnSe quantum wells, which
differ considerably in exciton and trion binding energy. The
normalized $X^{-}$ binding energy is found to be nearly
independent of electron-to-hole mass ratio for any quantum well
heterostructure with conventional parameters. Its dependence on
quantum well width follows an universal curve. The curve is
described by a simple phenomenological equation.
\end{abstract}


\maketitle


\section{\label{sec0} INTRODUCTION }

The first consideration of atomic-like three-body system is
regarded to Bethe, by whom the attention to the hydrogen ion
$H^{-}$ has been attracted as early as 1929 \cite{ref1}. The
existence of negatively (\textit{eeh}) and positively
(\textit{ehh}) charged excitons (trions) in semiconductors, being
analog of the hydrogen ions, was predicted by Lampert in 1958
\cite{ref2}. The investigation of three-body complexes has a
fundamental importance, particularly in semiconductors, where
there is a possibility to vary parameters in a wide spectrum.
However, the experimental observation of trions in bulk
semiconductors is rather difficult due to their small binding
energies.

The interest to experiment and theory of trions has grown due to
the progress in the semiconductor heterostructure fabrication.
Theoretical calculations performed at the end of the 1980s
\cite{ref3} predicted a considerable (up to tenfold) increase of
the trion binding energy in quantum well heterostructures compared
with bulk semiconductors. The first experimental observation of
negatively charged excitons ($X^{-}$) has been reported for
CdTe-based quantum wells (QWs) by K.~Kheng \textit{et al} in 1993
\cite{ref4}. The trions have also been observed in QWs based on
GaAs and ZnSe semiconductors \cite{ref5,ref6,ref7}. Nowadays, a
large number of experimental data on $X^{-}$ trion are available
for various types of heterostructures with different parameters.

The main characteristic of the negatively (or positively) charged
exciton is its binding energy, i.e. the energy required to
separate the trion in a neutral exciton and an unbound electron
(hole). The variation of the binding energy of $X^{-}$ trion
\cite{ref8,ref9,ref10,ref11} and, the similar system, $D^{-}$
center \cite{ref12,ref13} with the QW width have been extensively
studied theoretically. But, most of these calculations are limited
to specific material systems. In order to achieve a better
agreement with experimental data the problem is treated with a
considerable number of fitting parameters. This makes it very
difficult to compare trion binding energies in heterostructures
based on different semiconductors, which differ in Coulomb
energies.

The aim of this paper is to present a simple universal model,
which allows to estimate the trion binding energy ($E_{B}^{T}$) in
any semiconductor QW. In the following we show that the plausible
value of the $E_{B}^{T}$ at arbitrary QW width can be obtained
using a simple trial wave function, which provides a vivid picture
of the trion structure. The similar approach we used in
Refs.~\onlinecite{ref14,ref15,ref16} for the analysis of the
singlet and triplet states of trions in ideal two-dimensional
quantum wells and for the calculation of the trion ground state in
heterostructures with spatially separated carriers.

In this paper we concentrate on the negatively charged exciton,
which is caused by the reliable set of experimental data
available. It is important to note, that, commonly, the effective
mass of a hole is larger than that of an electron. So the $X^{-}$
is constructed of one heavy particle only, which simplifies the
theoretical consideration. The negatively charged trion can be
analyzed with the infinitely heavy hole centered at the QW, while
the electron-to-hole mass ratio, $\sigma = m_e / m_h$ , typically
being in the range of  $0.01 < \sigma < 1$, taken as a
perturbation parameter.

In section~\ref{sec1}, the experimental data for the trion binding
energy in heterostructures of different material systems are
summarized and discussed. In section~\ref{sec2}, a simple model of
the trion with a heavy hole in an ideal QW is proposed and the
binding energy dependence versus the effective well width is
variationally calculated. The mass ratio dependencies of the
exciton and the trion binding energies in the quantum well are
considered in section~\ref{sec3}. In section~\ref{sec4}, the
corrections to the trion binding energy due to QW imperfections
are discussed.


\section{\label{sec1} EXPERIMENTAL RESULTS}

\begin{table*}
\caption{\label{tab}The original experimental data collected for various
semiconductor materials. Note, that for correct comparison the binding energy
of "isolated" trion, which is unperturbed by interaction with two-dimensional
electron gas, must be taken into account \cite{ref31,ref18}. Therefore, we
either select the data for undoped QWs or extrapolate the binding energies in
doped structures to the low-concentration limit. In the latter case the initial
values are given in brackets. }
\begin{ruledtabular}
\begin{tabular}{c|c|ccccccccccc}
ZnSe   & $L_z$, {\AA} & 29 & 48 & 50 & 64 & 67 & 80 & 95 & 190 & 200 & & \\
       & $E_{B}^{T}$, meV & $8.9^{a}$
       & $6.6^{a}$ & $5.8^{a}$
       & $5.2^{a}$ & $5.3^{a}$ & $4.4^{a}$ & $4.0^{a}$
       & $1.4^{a}$ & $2.5^{b}$ & & \\
\hline
CdTe   & $L_z$, {\AA} & 38 & 50 & 55 & 80 & 100 & 120 & 150 & 260 & 400 & 500 & 600 \\
       & $E_{B}^{T}$, meV & $4.4^{c}$
       & $3.5^{c}$ & $3.4^{c}$
       & $2.9^{c}$ & $2.1^{d}$ & $2.5^{e}$ & $2.2^{c}$
       & $1.8^{c}$ & $1.3^{f}$ & $1.2^{l}$ & $1.1^{f}$\\
       & & & & & & (2.6) & & & & (1.8) & & (1.5) \\
\hline
GaAs   & $L_z$, {\AA} & 80 & 100 & 200 & 220 & 250 & 300 & & & & & \\
       & $E_{B}^{T}$, meV & $2.1^{g}$
       & $2.1^{h}$ & $1.15^{i}$
       & $1.1^{j}$ & $0.8^{k}$ & $0.9^{j}$ & & & & & \\
\end{tabular}
\end{ruledtabular}
\footnotetext{$^{a}$Reference~\onlinecite{ref18},
$^{b}$Reference~\onlinecite{ref24}, $^{c}$Reference~\onlinecite{ref19},
$^{d}$Reference~\onlinecite{ref31}, $^{e}$Reference~\onlinecite{ref25},
$^{f}$Reference~\onlinecite{ref26}, $^{g}$Reference~\onlinecite{ref27},
$^{h}$Reference~\onlinecite{ref28}, $^{i}$Reference~\onlinecite{ref29},
$^{j}$Reference~\onlinecite{ref30}, $^{k}$Reference~\onlinecite{ref17},
$^{l}$unpublished.}
\end{table*}

Charged excitons in various heterostructures have been extensively
studied during the last decade. Experimental data of the $X^{-}$
trion binding energy ($E_{B}^{T}$) for ZnSe, CdTe and GaAs quantum
wells of various widths ($L_z$) are collected in Table~\ref{tab}
and illustrated in Fig.~\ref{fig1}a. One can clearly follow the
increase of the trion binding energy by decreasing well width.

In spite of the variety of the experimental data, similarities
between the trions in different semiconductors are expected.
Indeed, considering the bulk trion in the frame of a simple model
of a Coulomb potential with the effective masses, one can take the
Bohr energy (i.e. exciton Rydberg), $Ry=\mu e^{4} / 2
\varepsilon^{2} \hbar^{2}$, and Bohr radius, $a_B=\hbar^{2}
\varepsilon / \mu e^{2}$ , as scaling parameters. Here $\mu=m_e
m_h/(m_e + m_h) $ is the reduced mass of the electron ($m_e$) and
the hole ($m_h$), $\varepsilon$ is the permittivity, and $e$ is
the electron charge. The binding energy of the trion normalized by
3D Rydberg is nearly independent of the electron-to-hole mass
ratio, $\sigma=m_e / m_h$ , and is $E_{B}^{T}\approx 0.055Ry$ for
most of the semiconductors studied \cite{ref3}. The same feature
is valid for the ideally 2D trion, i.e. for the trion being
strongly localized in the growth direction with the localization
length much smaller than $a_B$. In this case the trion binding
energy is $E_{B}^{T}\approx 0.48Ry$ and also shows no dependence
on the mass ratio ($\sigma$) and on the semiconductor material
\cite{ref3}. Therefore, it is rather natural to expect that the
$E_{B}^{T}$ does not strongly depend on $\sigma$ in the case of
finite width of a quantum well.

In order to compare experimental data for different materials collected in
Table~\ref{tab}, we plotted them in Bohr units, $E_{B}^{T}/Ry$ against
$L_z/a_B$, as shown in Fig.~\ref{fig1}b. The following values of 3D exciton
Rydberg $Ry=$ 4.2, 10, 20~meV and 3D exciton Bohr radius $a_B=$ 140, 67,
40~{\AA} were taken for GaAs, CdTe and ZnSe respectively \cite{ref32}. It can
be seen that, in these units, all dependences can be well approximated by one
universal curve. For example, a plausible estimation of the trion binding
energy in quantum wells of a thickness more than $a_B$ and less than $10a_B$
can be obtained with the simple fitting equation (shown in Fig.~\ref{fig1}b by
a solid line):
\begin{equation}
\frac{E_{B}^{T}}{Ry} \approx \frac{1}{3\sqrt{\frac{L_z}{a_B}}} \,.
\label{eq1}
\end{equation}
It is the simplest fitting equation found to well approximate the
experimental data. Of course, it cannot be used at the limiting
cases $L_z\rightarrow 0$ and $L_z\rightarrow \infty$. Nevertheless
it gives plausible estimation of the trion binding energy for the
wide range of $L_z$ and can be very useful due to its simplicity.

The fact that the experimental results for different
semiconductors coincide with the universal curve is remarkable. It
signifies that the trion binding energy for each effective width
of a QW is mainly scaled with Bohr units, and the influence of all
other parameters, i.e. electron-to-hole mass ratio or band
offsets, is rather weak.


\section{\label{sec2} TRION WITH INFINITELY HEAVY HOLE}

\begin{figure}
\includegraphics[width=.4\textwidth]{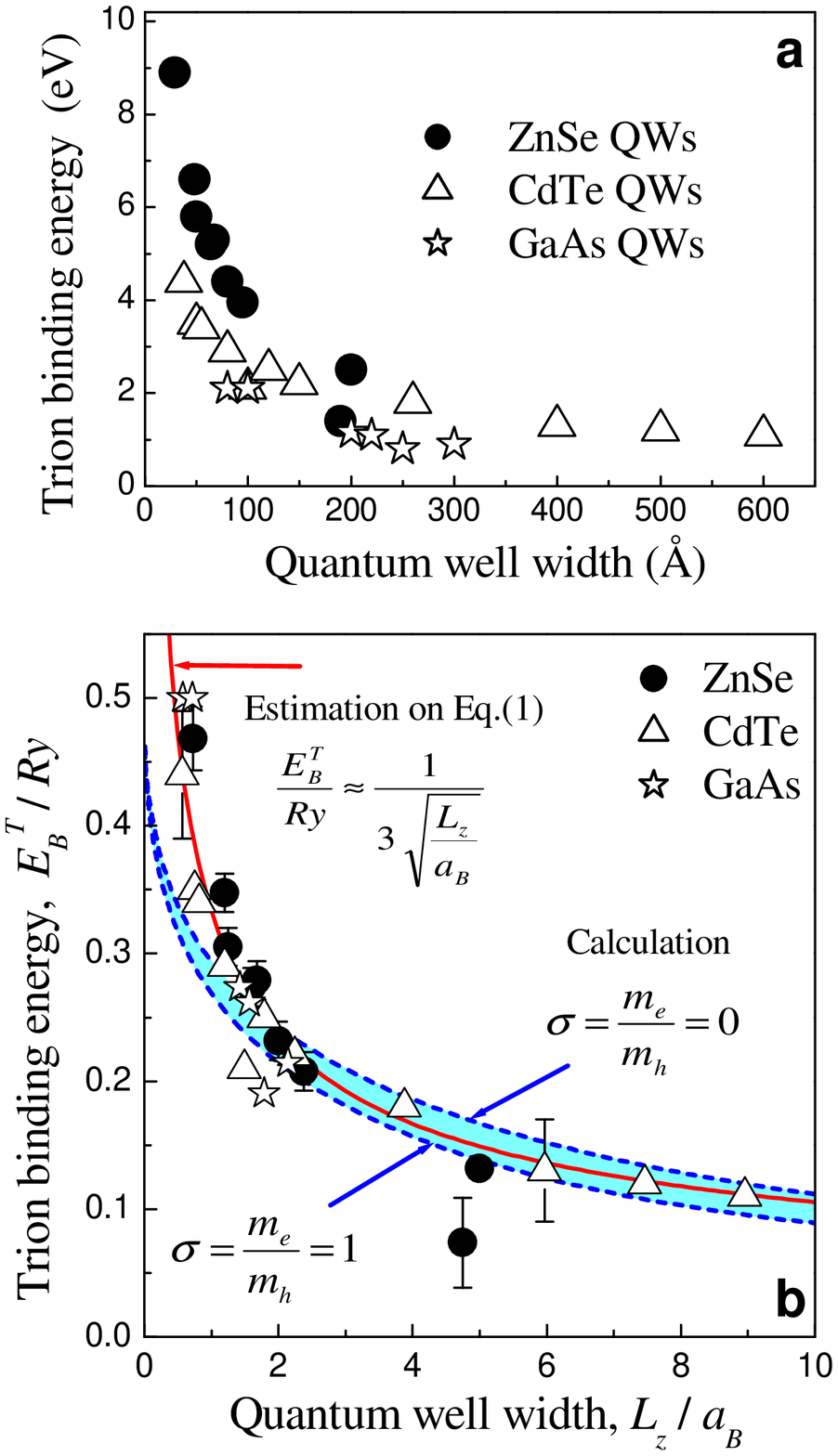}
\caption{\label{fig1} The $X^{-}$ trion binding energy $E_{B}^{T}$
versus the QW width  $L_z$ plotted for different semiconductors:
ZnSe by circles, CdTe by triangles, and GaAs by stars. These data
are also collected in Table~\ref{tab}. \\
(a) The experimental dependences are plotted in natural units,
i.e. energy and length being expressed in [eV] and [{\AA}],
respectively. \\
(b) The experimental dependences are plotted in 3D exciton units.
The solid line is estimation~(\ref{eq1}). The fill area represents
a scattering of calculated dependences due to the mass ratio
$\sigma$, obtained by variational method. It is confined by two
extreme cases with $\sigma =1$ and  $\sigma =0$. }
\end{figure}

In the previous section we showed that the experimental data of
the trion binding energy versus the QW width are well approximated
by the universal curve for all semiconductors, if the length and
the energy scales are expressed in exciton units. In this section
we obtain the universal dependence analytically. The requirement
of the universality greatly simplifies the task, because one can
leave only those parameters of the system, which can be directly
expressed in exciton energies and quantum well widths. Moreover,
in the next sections we show that the influence of other
parameters has no much effect on the universal curve obtained in
this section.

The simplification is the following. The trion is considered as a
three-body Coulomb system, using the effective mass approximation.
The reduced mass ($\mu$) and the permittivity ($\varepsilon$) are
supposed to be isotropic and identically in the quantum well and
in the barriers. The real potential of the quantum well is
replaced by an ideal one with infinite barriers. The hole is taken
to be much heavier than the electron, so the mass ratio $\sigma =
m_e / m_h$ is zero. In this case of only one heavy particle in the
system, namely the hole, it occupies the center of the quantum
well, where the adiabatic potential of the electrons reaches a
minimum. It reduces the number of independent coordinates in the
trion problem from 6 (the in-plane center-mass motion and the
total angular momentum along the growth direction are separated)
to 5. The number of the terms in the Hamiltonian decreases as
well.

The Schr\"{o}dinger equation for the trion in such a case is
(hereafter we use 3D exciton units for the length and the energy):
\begin{eqnarray}
[ -{\Delta}_{r_{1}} - {\Delta}_{r_{2}} + (
V_C(\mathbf{r_{1}},\mathbf{r_{2}}) + V^{QW}(z_{1}) +
\nonumber\\
V^{QW}(z_{2})) + E_{B}^{X} + E_{B}^{T} - 2E_{e}^{QW} ] {\Psi}_T
(\mathbf{r_{1}},\mathbf{r_{2}})=0 \,.  \label{eq2}
\end{eqnarray}
Here $\mathbf{r_{1}}$ and $\mathbf{r_{2}}$ are three-dimensional
vectors connecting the hole with the electrons, $z_{1}$ and
$z_{2}$ are their projections on the growth direction.
$V_C(\mathbf{r_{1}},\mathbf{r_{2}}) = 2(1/R - 1/r_{1} - 1/r_{2})$
is a Coulomb potential of the system, where
$R=|\mathbf{r_{1}}-\mathbf{r_{2}}|$ is the distance between the
electrons.  $E_{B}^{X}$ and $E_{B}^{T}$ are the exciton and trion
binding energies. $V^{QW}(z)$ is the quantum well potential, it is
0 if $|z|<L_z / 2$ and $+\infty$ otherwise. $E_{e}^{QW}$ is the
quantization energy of the free electron in the ground state of
the quantum well:
\begin{equation}
E_{e}^{QW} = \frac{1}{1+\sigma} \frac{\pi^{2}}{{L_z}^{2}}
\,\,\,\,\,\,\,\,\, \left(\mathrm{if} \,\,\, \sigma = 0, \,\,\,
E_{e}^{QW} =\frac{\pi^{2}}{{L_z}^{2}} \right) \,. \label{eq3}
\end{equation}
The equation~(\ref{eq2}) is solved by a variational method, when
the energy in the ground state, calculated by a trial function
with variational parameters, is minimized. The most critical
point, effecting the accuracy of calculations, is the proper
choice of the trial function, which should be simple and close to
the real wave function. In order to obtain the trion binding
energy one has to find the trion energy and subtract it from that
of the exciton. Therefore, to minimize the mistake in the
$E_{B}^{T}$ calculation, the exciton energy should be calculated
in the same manner as the trion one. Consequently, the trion
function should be based on the exciton function, transforming to
the latter when one of the electrons is removed.

The simplest trial function for the exciton with only one
variational parameter ($a$), which gives plausible results for the
exciton binding energy in the whole range of the quantum well
widths, is:
\begin{equation}
{\Psi}_X(\mathbf{r}) = A \,\, \exp(-a \, r) \, Z_0(z,L_z) \,,
\label{eq4}
\end{equation}
Here $\mathbf{r_{1}}$ is 3D vector connecting the hole and
electron, and $z$ is its projection on the growth direction. $A$,
here and after, is a normalization factor of the corresponding
wave function. The last multiplier, $Z_0(z,L_z)$, provides the
additional localization of the electron in the growth direction
due to the quantum well potential. It should be stressed that
in-plane and in-growth motion of the electron in the function (4)
is not separated, and the influence of the electron-hole
interaction on the in-growth electron motion is substantially
taken into account by the exponential multiplier. So, in contrast
to the adiabatic case, where the excited states of the quantum
well have to be included (see, for example, [10]), we can take
$Z_0(z,L_z)$ as a function of the ground state:
\begin{eqnarray}
Z_0(z,L_z) = \sqrt{\frac{2}{L_z}}\cos(\pi \frac{z}{L_z}), \,\,\,\,
\mathrm{for} \,\, |z| \leq L_z/2,
\nonumber\\
Z_0(z,L_z) = 0,
\,\,\,\,\,\,\,\,\,\,\,\,\,\,\,\,\,\,\,\,\,\,\,\,\,\,\,\,\,\,\,\,\,\,\,\,\,\,\,\,\,
\mathrm{for} \,\, |z| > L_z/2. \, \label{eq5}
\end{eqnarray}
Here $\mathbf{r}$ is 3D vector connecting the hole and electron,
and $z$ is its projection on the growth direction. It is easy to
see that function~(\ref{eq4}) turns into the exact wave function
of the exciton in both limiting cases of an ideal 2D quantum well
($L_z \rightarrow 0$) and a 3D bulk semiconductor ($L_z
\rightarrow \infty$). Besides the simplicity, the
function~(\ref{eq4}) has one more benefit. It can be shown that
the full kinetic energy of the electron in the case of any
arbitrary quantum well potential $V^{QW}(z)$ is:
\begin{eqnarray}
E_e^{kin} = \left\langle -{\Delta}_r \right\rangle =
\,\,\,\,\,\,\,\,\,\,\,\,\,\,\,\,\,\,\,\,\,\,\,\,\,\,\,
\,\,\,\,\,\,\,\,\,\,\,\,\,\,\,\,\,\,\,\,\,\,\,\,\,\,\,
\,\,\,\,\,\,\,\,\,\,\,\,\,\,\,\,\,\,\,\,\,\,\,\,\,\,\,
\nonumber\\
\left\langle \frac{\partial}{\partial r} {\Psi}_X(r,z) \mid
\frac{\partial}{\partial r} {\Psi}_X(r,z) \right\rangle +
E_e^{QW}-
\nonumber\\
\left\langle {\Psi}_X | V^{QW} | {\Psi}_X \right\rangle = a^2 +
E_e^{QW} - \left\langle V^{QW} \right\rangle \,. \label{eq6}
\end{eqnarray}
Consequently, the quantization energy $E_e^{QW}$  and the mean
value of the quantum well potential  $V^{QW}$ in the
Schr\"{o}dinger equation can be eliminated analytically. In this
case the mistakes arising from the application of numerically
methods are avoided, which considerably simplifies the
calculations. Consequently, the binding energy of the exciton
($E_{B}^{X}$) can be estimated by the formula:
\begin{equation}
E_{B}^{X} = -{\min}_a (a^2 - \left\langle V_C (r) \right\rangle )
\,. \label{eq7}
\end{equation}
Here $V_C (r) = -2/r$ is a Coulomb potential between the electron
and the hole. It should be noted, that the parameters of the
quantum well are included in the mean value of the Coulomb
potential through the last multiplier of the function~(\ref{eq4}).
The first term in Eq.~(\ref{eq7}) also has a slight dependence on
the quantum well structure if the exponent in the
function~(\ref{eq4}) is replaced by any other radial function. The
equalities similar to (\ref{eq6}-\ref{eq7}) are valid for all
trial functions considered below.

The simplest trion function, based on the exciton
function~(\ref{eq4}), is the 3-parameter Chandrasekhar-like one
\cite{ref23}:
\begin{eqnarray}
{\Psi}_T (\mathbf{r_{1}} , \mathbf{r_{2}}) = A ( \exp (- a_{1}
r_{1} - a_{2} r_{2}) +  \nonumber\\
\exp (- a_{2} r_{1} - a_{1} r_{2}) ) \, (1 + c R) Z_0 (z_{1}, L_z)
Z_0 (z_{2}, L_z) \,.  \label{eq8}
\end{eqnarray}
Here $a_{1}$, $a_{2}$, and $c$ are variational parameters.
Analogously to the function~(\ref{eq4}), this function transforms
into the appropriate Chandrasekhar's one in the limiting cases of
two and three dimensions:
\begin{eqnarray}
{\Psi}_T (\mathbf{r_{1}} , \mathbf{r_{2}}) = A ( \exp (- a_{1}
r_{1} - a_{2} r_{2}) + \,\,\,\,\,\,\,\,\,\,\,\,\,\,\,\,\,\,\,\,\, \nonumber\\
\,\,\,\,\,\,\,\,\,\,\,\,\,\,\,\,\,\,\,\,\,\, \exp (- a_{2} r_{1} -
a_{1} r_{2}) ) \, (1 + c R) \,. \label{eq9}
\end{eqnarray}
The relative mistake in the trion binding energy obtained with
function~(\ref{eq8}) is known to be less than 10\% both in the 2D
and 3D cases. Therefore, we can expect that the estimations, given
by (\ref{eq8}) even in the intermediate cases of the finite-width
quantum wells, are also not far from the exact values.

The calculated trion binding energy versus the quantum well width
within the described approach is shown in the Fig.~\ref{fig1}b
(the dashed line pointed by $\sigma = 0$). The calculation is in
very good agreement with the experimental data for wide quantum
wells ($L_z \geq 2 a_B$). However, in narrow ($L_z < 2 a_B$)
quantum wells the discrepancy becomes considerable. The possible
reasons for this will be discussed in section~\ref{sec4}.


\section{\label{sec3} MASS RATIO EFFECT IN THE EXCITON AND TRION IN QW}

It is known that the binding energy of $X^{-}$ trion, expressed in
Bohr units, is nearly independent of electron-to-hole mass ratio
$\sigma$ both for an ideal 2D quantum wells and for 3D bulk
semiconductor \cite{ref3}. It is rather natural to expect that in
the quantum well of arbitrary width the $X^{-}$  trion binding
energy to be weakly dependent on $\sigma$. The results of
calculation of the trion binding energy versus the mass ratio for
a 250 {\AA}-wide GaAs QW also confirm this assumption
\cite{ref17}.

The correction to the trion binding energy due to nonzero mass
ratio is supposed to be small. Therefore in what follows we will
consider it as a perturbation and calculate in the adiabatic
approximation. To simplify this calculation we start by analyzing
the exciton and then expand the results to the trion case. In the
case of the exciton with a very heavy hole, the particle wave
functions are separated and the adiabatic approximation is
applicable (for narrow quantum wells the influence of the in-plane
electron motion on the in-growth hole localization is fully
analyzed in \cite{ref34}). The Schr\"{o}dinger equation for the
hole motion in the growth direction is (in 3D exciton units):
\begin{equation}
\left( -\frac{\sigma}{1+\sigma} \frac{{\partial}^2}{\partial z^2}
+ \left( V_e^{adiab}(z) + E_{B}^{X} - E_h^{QW} \right) \right) Z_h
(z) = 0  \,. \label{eq10}
\end{equation}
Here $Z_h(z)$ is the wave function of the hole in growth
direction. $V_e^{adiab}(z)$ is a sum of the averaged Coulomb
potential of the electron and the quantum well. $E_h^{QW}$ is a
quantization energy of the free hole in the ground state of the
quantum well:
\begin{equation}
E_h^{QW} = \frac{\sigma}{1+\sigma} \frac{{\pi}^2}{{L_z}^2} \,.
\label{eq11}
\end{equation}
As mentioned in the previous section, the quantum well potential
is taken to be ideal, with infinite barriers. It is easy to show
that the hole with infinitely heavy mass ($\sigma=0$) is located
in the minimum of the adiabatic potential ($z=0$). The binding
energy of the exciton in such case is:
\begin{equation}
E_{B}^{X} = -V_{e}^{adiab} (0) \,. \label{eq12}
\end{equation}
As the mass ratio increases, the binding energy of the exciton
decreases because, by the definition:
\begin{eqnarray}
\left\langle Z_h | - \frac{\sigma}{1+\sigma}
\frac{{\partial}^2}{\partial z^2} | Z_h \right\rangle \geq
E_h^{QW} \,\,\, \nonumber\\
\left\langle Z_h | V_e^{adiab} (z) | Z_h \right\rangle \geq
V_e^{adiab} (0) \,. \label{eq13}
\end{eqnarray}
Qualitatively, if the mass of the hole becomes smaller, its
localization along z-direction increases until it achieves the
width of the QW, and then stays unchanged. Therefore, the main
factor defining the evolution of the exciton binding energy with
the mass ratio is the hole localization in the growth direction
due to the Coulomb attraction of the electron. The simplest wave
function taking this into account is:
\begin{equation}
{\Psi}_X(r, z_e, z_h) = A \,\, \exp(-a \, r) \, Z_0(z_e,L_z) \,
Z_0(z_h,(bL_z)) \,. \label{eq14}
\end{equation}
Here $r$ is the 3D distance between the particles, $a$ is the
reciprocal radius of the exciton, $b\in [0,1]$  is the degree of
hole localization. The value $b=1$ means the function of the hole
in the growth direction is nearly the same as that of the
electron. The opposite case, $b=0$, signifies that the hole is
strongly localized in the center of the well corresponding to the
case of an infinitely heavy hole.

\begin{figure}
\includegraphics[width=.4\textwidth]{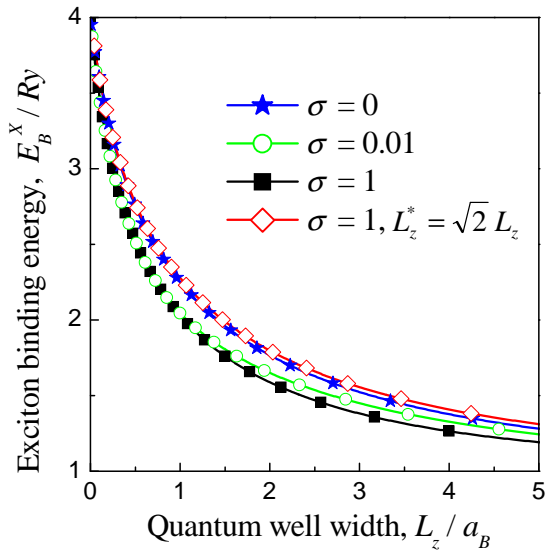}
\caption{\label{fig2} The exciton binding energy $E_{B}^{X}$
calculated versus the quantum well width $L_z$ for different
values of the mass ratio $\sigma = m_e / m_h$. The curves with
$\sigma =1$ (squares) and $\sigma =0.01$ (circles) are nearly
coincide. The curve with $\sigma =0$ (stars) is precisely
approximated by the rescaled dependence with $\sigma = 1$
(diamonds), the coefficient being $\sqrt{2}$. }
\end{figure}

The dependences $E_{B}^{X}(L_z)$ are calculated by variational
method with the trial function~(\ref{eq14}) for few values of mass
ratio $\sigma = m_e / m_h$ (see Fig.~\ref{fig2}). The difference
in energy between even the extreme curves $\sigma =0$ and $\sigma
=1$ is rather small ($<$10\%) for all values of $L_z$. Moreover,
it can be noticed that the curve corresponding to $\sigma =0.01$
is closer to the curve for $\sigma =1$ than $\sigma =0$. For
example, the curves  $\sigma =0.1$ and $\sigma =1$ would not be
distinguishable in the scale of the figure. It means, that the
exciton binding energy does not depend on the mass ratio for
$\sigma >0.1$ and has an extremely weak dependence if $\sigma \in
[0.1,0.01]$. A considerable increase of $E_{B}^{X}$ takes place
only if the hole is unrealistic heavy ($\sigma <0.01$).
Consequently, we can neglect the variation of the binding energy
with the mass ratio for all experimental values of the latter.

\begin{figure}
\includegraphics[width=.4\textwidth]{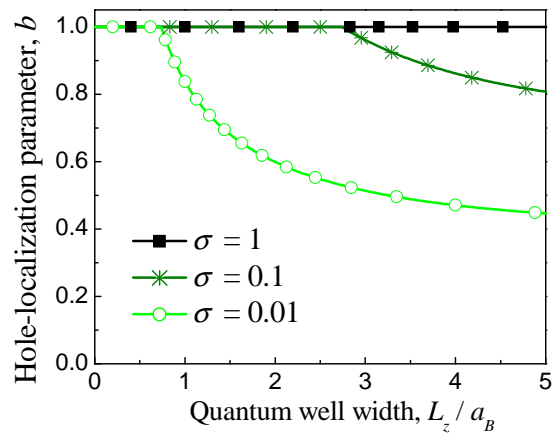}
\caption{\label{fig3} The parameter of the hole in-growth
localization ($b$) versus the quantum well width  $L_z$ for
different values of the mass ratio $\sigma = m_e / m_h=$  $0.01$
(circles), $0.1$ (stars), and $1$ (squares). }
\end{figure}

This fact is easy to understand, if we consider the degree of the
hole localization ($b$) in such cases (see Fig.~\ref{fig3}). It
can be seen that even if the hole is rather heavy (for example,
$\sigma = 0.1$) the value of the hole localization in the growth
direction is nearly $b \sim 1$ for all $L_z$, whereas the limiting
value $\sigma =0$ corresponds to $b=0$. It means that the exciton
wave function~(\ref{eq14}) is nearly independent of the mass ratio
for $\sigma >0.1$. Consequently, due to Eq.~(\ref{eq7}), which is
valid for function~(\ref{eq14}) as well, $E_{B}^{X}$ appears also
to have no dependence on $\sigma$ in the same value range.

Note, the curve with  $\sigma =0$ in Fig.~\ref{fig2} can easily be
obtained from the curve with $\sigma =1$ if the abscissa is
multiplied by a coefficient $\sqrt{2}$. A small discrepancy
between these curves takes place only for wide quantum wells, but
even there, it is nearly negligible. This is a consequence of the
fact, that the function~(\ref{eq14}) involves in an explicit form
the electron and hole z-coordinates ($z_e$ and $z_h$) only through
the functions $Z_0$. Indeed, for the case of:
\begin{equation}
Z_0(z,L) = \frac{1}{\sqrt{\sqrt{\pi}L}} \exp \left( -
\frac{z^2}{2L^2} \right) \,, \label{eq15}
\end{equation}
it can exactly be shown, that the binding energy of the exciton,
given by Eq.~(\ref{eq7}), is the same both for $b\equiv 1$,  $L
\equiv L_z$ and $b\equiv 0$,  $L \equiv \sqrt{2}L_z$:
\begin{equation}
E_{B}^{X}(L_z,\sigma =1) \equiv E_{B}^{X}(\sqrt{2}L_z,\sigma =0)
\,. \label{eq16}
\end{equation}
The equality~(\ref{eq16}) is valid even if the exponent in the
exciton function~(\ref{eq14}) is replaced by any other radial
function. However, if the function $Z_0$ differs from a Gaussian
function, the equality~(\ref{eq16}) becomes not valid.
Nevertheless, as can be seen in Fig.~\ref{fig2}, it produces a
good estimation of the binding energy of the light-hole exciton
(i.e. with  $\sigma =1$) for a wide range of quantum well width
values.

The obtained results for the exciton can be extended to the trion.
It is known that the binding energy of the second electron in the
trion is much smaller ($\sim 10$ times) than that of the first
one. Therefore, the negatively charged trion, containing only one
heavy particle, can be considered in a rather crude model as an
electron bound to an unperturbed exciton via some effective
central potential $V_X^{eff}(r)$. Here $r$ is the distance between
the electron and the center of mass of the exciton, which is
assumed to be unperturbed. In that way, the problem of the trion
becomes very similar to the exciton one considered in this section
earlier. Therefore, one can suppose that the only effect which
causes an alteration of $E_{B}^{T}(L_z, \sigma)$ with $\sigma$ is
the increase of the exciton localization in the growth direction
due to the interaction with the additional electron. By analogy to
Eq.~(\ref{eq16}), the dependence of the trion binding energy with
a mass ratio $\sigma =1$ can be obtained via rescaling the curve
with $\sigma =0$:
\begin{equation}
E_{B}^{T}(L_z,1) \approx E_{B}^{T}(\sqrt{2}L_z,0) \,, \label{eq17}
\end{equation}
where the latter is known from the previous section. Obviously,
all possible dependences on the QW width of the trion binding
energy are confined by these two extreme cases with $\sigma =1$
and $\sigma =0$, as are presented in Fig.~\ref{fig1}b by the
filled area. The obtained scattering of the binding energy is less
than $20 \%$, which is even smaller than the experimental data
dispersion. Moreover, as for the exciton binding energy, the trion
$E_{B}^{T}$ is expected to be about the same for $\sigma >0.01$,
allowing to take $\sigma =1$ for any QW with realistic parameters.
All these arguments prove the thesis, that the binding energy of
$X^{-}$ trion is nearly independent of the mass ratio in most
quantum well heterostructures.

It is worth to note that our considerations taken for GaAs-based
QWs generally represent the results of previous numerical
calculations \cite{ref8,ref9,ref10,ref11}. However, in contrast to
them, in the present paper the theoretical results have been
obtained with the use of only three fitting parameters, and an
agreement is achieved for various semiconductor systems (i.e. for
CdTe and ZnSe in additional to GaAs).


\section{\label{sec4} CORRECTIONS TO THE TRION BINDING ENERGY}

In the previous sections the corrections to the trion binding
energy, appearing from those parameters of QWs, which cannot be
expressed in Bohr units, are neglected for simplification. The
relatively small dispersion of the experimental data for
$E_{B}^{T}$ in different semiconductors allows us to conclude that
the scale of these corrections is at most in the range of 20\%
(see Fig.~\ref{fig1}b). In this section we discuss possible
corrections and evaluate their input in the trion binding energy.

\textit{Correction due to lateral localization.} The largest
difference between the calculation and the experiment is observed
for narrow QWs (Fig.~\ref{fig1}b). The reasonable explanation of
this fact is that the trion binding energy increases due to
in-plane localization at one-monolayer fluctuations of the QW
width (see, for example, \cite{ref33}). As the QW becomes
narrower, these fluctuations, forming lateral islands, become more
important. The energy alteration due to this effect strongly
depends on the effective size of the islands, which are controlled
by growth conditions and expected to be individual for each
sample. The formation of lateral islands is confirmed by the
broadening of the trion line in optical spectra being observed
experimentally \cite{ref18}. However, it is not clear up to now,
why the results for different semiconductors are so similar.

\textit{Polaron correction.} It has been assumed that the
interaction between the particles is of Coulomb type, and the
effective mass approximation has been used. However, the polaron
correction is known to cause a considerable increase of the
binding energy of $D^{-}$ center in narrow quantum wells due to
reducing the distance between the electrons \cite{ref12}. It might
also explain the discrepancy of the experimental and the
theoretical data for narrow quantum wells (see Fig.~\ref{fig1}b).
The polaron correction was also used in fitting the experimental
data for the trion binding energy reported in
Ref.~\onlinecite{ref9}. Nevertheless, it must be stressed, that
the polaron correction for $X^{-}$ trion is expected to be nearly
the same as for $D^{-}$ center. But the correction for the $D^{-}$
binding energy quickly saturates in wide quantum wells and remains
quite significant there, whereas the discrepancy between the
theory and experiment in Fig.~\ref{fig1}b is negligible small for
wide quantum wells.

\textit{Corrections due to anisotropy.} The next simplification,
the reduced mass ($\mu$) and the permittivity ($\varepsilon$) have
been taken to be isotropic and to have the same values in the
quantum well and in barriers. It allows us to take the Bohr energy
and the Bohr radius of the QW material as a system scale. It is
clear that the role of the mass and permittivity anisotropy
diminishes in the limit of two dimensions. Therefore, if the
anisotropy is small enough to considerably change the bulk trion
energy, then it can be neglected in finite quantum wells as well.
The discontinuity of the permittivity across the interfaces causes
image charges in the barrier areas \cite{ref20,ref21}. For the
typical situation of smaller   in the barriers compared to the
value in the wells, the Coulomb interaction (i.e. binding energy)
between particles is effectively increased. For example, the
relative increase of the exciton binding energy in narrow GaAs
quantum wells due to this effect is about 10-20\% \cite{ref22}.
This value is comparable with the error of our estimations and
cannot qualitatively change the results.

\textit{Corrections caused by finite barriers.} The main effect of
the quantum well potential is the localization of carriers in the
structure growth direction. The binding energy of the trion
depends more on the localization degree, than on the real shape of
the quantum well potential. The simplest way to take it into
account is to consider the real quantum well as an ideal one with
infinite barriers and different effective widths for electrons
$(L_e)$ and holes $(L_h)$, which is caused indeed by penetration
of their wave functions in the barriers. These quantities should
be treated as phenomenological parameters and, with a reasonable
accuracy, can be taken so that the mean-square deviation of the
particles in the growth direction remains the same in the ideal
quantum well as in the real one. As has been shown above, a small
relative inaccuracy in the effective well width does not lead to a
considerable change of the binding energy. For example, if the
width is taken to be 10\% larger the binding energy decreases at
most by 4\%, which is quite small compared with the uncertainty of
experimental data. In wide quantum wells $L_h$ is nearly equal to
$L_e$, and one can take $L_e \approx L_h \approx L_z$. In narrow
QWs, these values can be considerably different $L_e \neq L_h$ due
to the tails of wave functions penetrating into the barrier, which
depend on the barrier specifics for the electron and the hole.

In most cases of narrow QWs $L_h < L_e$, which is easy to
consider. Indeed, as $L_h$ decreases, the electron-hole
interaction becomes stronger and the trion binding energy
increases. The case $L_h \rightarrow 0$ is very similar to the
case when $\sigma =0$, considered in the section~\ref{sec2}.
Therefore, in the frames of model considered, the simple
estimation can be obtained:
\begin{eqnarray}
E_{B}^{T}(L_e,L_e,1) \leq E_{B}^{T}(L_e,L_e,\sigma) \leq
\nonumber\\
E_{B}^{T}(L_e,L_h,\sigma) \leq E_{B}^{T}(L_e,L_h,0) =
E_{B}^{T}(L_e,L_e,0) \,, \label{eq18}
\end{eqnarray}
where the relative difference between the boundaries does not
exceed 20\% (Fig.~\ref{fig1}b). Moreover, the trion binding energy
can be estimated more accurately. If $\sigma$ is not small enough
to cause a considerable localization of the hole in the growth
direction ($\sigma > 0.01$), one can take:
\begin{eqnarray}
E_{B}^{T}(L_e,L_h,\sigma) \approx E_{B}^{T}(L_e,L_h,1) \approx
E_{B}^{T}(L_z^{*},L_z^{*},0) \,, \nonumber\\
\mathrm{where} \,\,\,\,\,\,\,\,\,  L_z^{*} = \sqrt{{L_e}^2 +
{L_h}^2} \,. \label{eq19}
\end{eqnarray}
In the same way, as it is shown for Eq.~(\ref{eq16}), the last
equality can be rigorously proven, if the ground state functions
of the electron and hole in the quantum well are Gaussian
functions~(\ref{eq15}). However, the estimation~(\ref{eq19}) can
still be used in very arbitrary quantum wells.

The opposite case, $L_h > L_e$, corresponding to a extremely
shallow potential for the hole, is much more complicated. In the
extreme case of $L_e \rightarrow 0$ and $L_h \rightarrow \infty$
the hole is localized in the field of electrons only. In this
limit, if $\sigma =0$, the trion binding energy is equal to that
of an ideally 2D case, because the hole is localized in the same
plane as the electron. However, if $\sigma$ grows, the zero-point
oscillations of the hole becomes important. In the case of the
exciton, the adiabatic potential, produced by the electron in
growth direction, is:
\begin{equation}
V_h^{adiab}(z) = - 4 + 16|z|, \,\,\,\,\,\,\,\,\,\, z \rightarrow
0\,. \label{eq20}
\end{equation}
Accordingly, it is easy to obtain, that the exciton binding energy
quickly decreases with the increase of $\sigma$:
\begin{equation}
\, E_{B}^{X}(\sigma) \approx 4 - 6.5 \, {\sigma}^{1/3},
\,\,\,\,\,\,\,\,\,\, \sigma \rightarrow 0\,. \label{eq21}
\end{equation}
The effect for the trion is even stronger. For example, if
$\sigma$ becomes $\sim 1$, the binding energy of the $X^{-}$ trion
decreases about one order of magnitude compared to $\sigma =0$.
This happens because, after averaging over the z-coordinate of the
hole, the electron-hole interaction becomes weaker than the
electron-electron one. Such situation is similar to the system
with spatially separated carriers, so the effective distance
between the carrier layers increases, as the mass of the hole
becomes smaller.

\textit{Corrections owing to a build-in electric field.} Special
attention should be drawn, if the quantum well is not symmetric,
and there is a spatial separation in the growth direction between
the carriers. In such a case, it should be taken into account that
the binding energy of the trion strongly reduces with the distance
between the electron and hole layers. For example, if the
splitting of the layers becomes $\sim a_B$, the binding energy of
the $X^{-}$ trion decreases more than one order of magnitude
\cite{ref16}. It is very possible, that the relatively low binding
energy observed in the wide ZnSe quantum well (see
Fig.~\ref{fig1}b) is caused by a build-in electric field or a
quantum well asymmetry leading to some spatial separation of the
carriers.

Thus, it has been shown, that for the most of the quantum wells,
the binding energy of the negatively charged trion can be easily
estimated by means of the universal model containing only few
parameters.

\section{\label{sec5} CONCLUSIONS}

The experimental values of the trion binding energy for various
semiconductor quantum wells, being represented in corresponding
exciton units, are found to be well approximated by an universal
function. The theoretical estimations confirm that in a simplified
Coulomb model the $X^{-}$  trion binding energy is nearly
independent of the electron-to-hole mass ratio at any value of
quantum well width. Consequently, for the sake of simplicity,
calculations of the trion binding energy can be performed with
infinite hole mass values. In narrow quantum wells the
experimental data cannot be explained in the frame of an idealized
model and additional factors should be involved in the
consideration.


\begin{acknowledgments}

This work was supported by the Deutche Forschungsgemeinschaft via
Sonderforschungsbereich 410, Russian Foundation for Basic Research
and the Federal Program on Support of Leading Scientific Schools.
One of the authors (R.A. Suris) appreciates the support of the
Alexander von Humboldt Foundation.

\end{acknowledgments}



\begin{thebibliography}{34}
\bibitem{ref32} Through the reference papers there is some dispersion in values of bulk exciton Rydberg
  and Bohr radius. Therefore we took the averaged values, which might differ from the real ones at the most
  of 15{\%}.
\end{thebibliography}
\end{document}